\documentstyle[prl,aps]{revtex}
\begin{document}

\title{Breakdown of the perturbative renormalization group for 
 $S\ge 1$ random antiferromagnetic spin chains}

\author{ \sc {A. Saguia, B. Boechat and M. A. Continentino}}
\address{Instituto de F\'{\i}sica - Universidade Federal Fluminense\\
24210-130, Niter\'oi, RJ - Brazil; \\}

\author{ \sc { O. \ F. \ de Alcantara Bonfim }} 
\address{Department of Physics, Reed College, Portland Oregon 97202, USA}

\date{\today}

\maketitle

\begin{abstract}

We investigate the application of a perturbative renormalization
group (RG) method to random antiferromagnetic Heisenberg chains with arbitrary
spin size. At zero temperature we observe that initial arbitrary probability
distributions develop a singularity at $J=0$, for all values of spin $S$. When
the RG method is extended to finite temperatures, without any additional
assumptions, we find anomalous results for $S\ge 1$. These results lead us to
conclude that the perturbative scheme is not adequate to study random chains
with $S\ge 1$. Therefore a random singlet phase in its more restrictive
definition is only assured for spin-1/2 chains.  


\end{abstract}

\vspace{1.0cm}

Low-dimensional random quantum magnetic systems have become an object of
increasing interest in recent years.
Despite the apparent simplicity, quantum spin chains  show a wealth of
physical properties which have attracted the attention of both theoretical and
experimental physicists. It is generally accepted that quantum
antiferromagnetic spin-1/2 chains in the presence of disorder exhibit, at low
temperatures, a random singlet (RS) phase characterized  by power
law behavior of thermodynamic quantities\cite{MDH,DFisher}. This has,  in
fact, been observed in several
experiments~\cite{JCScott,YEndoh,LNBula,GTheo,JCarlos}. Theoretically, the  new
phase was revealed by a  real space renormalization group (MDH)
approach\cite{MDH}.  These  results have motivated many authors to  study the
physical properties of other  random antiferromagnetic quantum (RAQ) chains
with $S\ge 1$. In this case there are additional new problems, as for example,
what happens to the Haldane gap of integer spins chains\cite{Haldane} in the
presence of disorder. In fact, the possibility of the  energy gap be
suppressed by  disorder, driving the system to a RS phase has become a very
attractive matter for investigation. In particular, the $S=1$ RAQ chains has
been exhaustively  studied by  several
methods\cite{bia1,bia2,Hyman,Jolicoeur,bia3,Nis,MC,Hida}. Many authors
have employed  extended versions of the  MDH scheme to
explore the  ground state properties of the model. In the course of these
investigations, some authors suggested  the existence of random singlet
phase in the spin-1 chain in the strong disorder
regime~\cite{bia1,bia2,Hyman,Jolicoeur,bia3}. On the other hand,   studies
based on exact diagonalization~\cite{Nis}, quantum Monte Carlo (MC)
simulation~\cite{MC} and  density matrix  renormalization group (DMRG)
techniques\cite{Hida} found that the   Haldane phase is quite robust against
randomness.

In view of these controversial results and the fact that most calculations for
spin-1 chains have been done at $T=0$, 
we were motivated to extend the MDH analysis to finite temperature.
Our goal was to generalize the MDH recursion relations to study the
thermodynamic behavior of chains with arbitrary spin size. 
We performed careful numerical procedures to iterate  our generalized MDH
equations at finite temperature. The anomalous results we have obtained for the
free energy and are presented in this communication  indicate that the MDH
approach breaks down for chains with $S\ge 1$. Therefore, we argue that it is
impossible to conclude on the existence or not of the RS phase  in the $S\ge 1$
RAQ chains,  as proposed in previous zero temperature
studies using the MDH method\cite{bia1,bia2,Hyman,Jolicoeur,bia3}.  Actually,
within the MDH approximation the existence of the RS phase  is assured only
for the spin-1/2 chains.  

 The random magnetic  systems  are described by the
Heisenberg  Hamiltonian:
\begin{equation} 
H = \sum_{i=1}^{L-1} J_{i}
{\vec{S}}_i .  {\vec{S}}_{i+1}, 
\label{eq1.1}
\end{equation}
where $J_{i} >0$ is the  nearest neighbor antiferromagnetic interaction
and  $\vec{S}_{i}$ are quantum spin  operators. The exchange couplings are
random variables distributed according to a given probability distribution
$P_J(J_i,\Omega )$, with a cutoff $\Omega$.  
The MDH method consists in eliminating
the pair of spins with 
the strongest  coupling ($J_2 = \Omega$) in the random chain  
by considering the interaction ($J_1$ and $J_3$) with the neighboring spins 
of this pair as a perturbation (see Fig.~\ref{cad}).
  
The Hamiltonian for the strongly coupled pair of spins ${\vec{S}}_2$ and 
${\vec{S}}_3$ is given by
\begin{equation}
H_{0}=J_{2}\vec{S_{2}}.\vec{S}_{3}.
\label{h0}
\end{equation}
These spins  are weakly coupled to the neighbors via
\begin{eqnarray}
H_{1}= J_{1}{\vec{S}}_{1}.{\vec{S}}_{2}
+J_{3}{\vec{S}}_{3}.{\vec{S}}_{4}.
\label{h1}
\end{eqnarray}
We calculate the free energy of the system formed by ${\vec{S}}_1$, 
${\vec{S}}_2$, ${\vec{S}}_3$ and ${\vec{S}}_4$ through perturbation
theory in second order of $H_{1}$ (see ref.\cite{MDH}).
After some simple but extensive  calculations we come with the results:
\begin{equation}
F^{\prime} = F_{0}-\frac{1}{3\Omega}[S(S+1)]^{2}(J^{2}_{1}+
J^{2}_{3})V_{s}(\beta\Omega)
\label{Fprime}
\end{equation}
with
\begin{equation}
F_{0}=-S(S+1)\Omega - \frac{1}{\beta}\ln {\sum^{2S}_{i=0} (2i+1)
\exp [-\frac{1}{2}i(i+1)\beta\Omega]}
\label{fos}
\end{equation}
and
\begin{equation}
J^{\prime} = \frac{2}{3}S(S+1)\frac{J_{1}J_{3}}{\Omega}W_{s}(\beta \Omega),
\label{Kprime}
\end{equation}
where $\beta = 1/k_{B}T$ and the functions  V and W are given by
\begin{equation}
V_{s}(y) = \frac{(2S+1)^{2}-\sum^{2S}_{i=0} (2i+1)
e^{-\frac{1}{2}i(i+1)y}[1-\frac{1}{2}i(i+1)y]}
{4S(S+1)\sum^{2S}_{i=0} (2i+1)e^{-\frac{1}{2}i(i+1)y}}\\
\label{Ve}
\end{equation}
\noindent and
\begin{equation}
W_{s}(y) = \frac{(2S+1)^{2}-\sum^{2S}_{i=0} (2i+1)
e^{-\frac{1}{2}i(i+1)y}[1+\frac{1}{2}i(i+1)y]}
{4S(S+1)\sum^{2S}_{i=0} (2i+1)e^{-\frac{1}{2}i(i+1)y}}.
\label{Wb}
\end{equation}
Equations~(\ref{Fprime})-(\ref{Wb}) are the generalization of the 
MDH equations for the thermodynamic properties of RAQ chains with
arbitrary spin size. For S=1/2 we recover the original MDH recursion
relations\cite{MDH}. The S=1 case has already been studied 
in the limit of very low temperatures\cite{bia2}.

When the decimation process is carried out for a spin-1/2 chain
the distribution $P_J(J_{i},\Omega )$, independently of the form of the
original distribution of antiferromagnetic bonds, rather quickly approaches a
fixed form singular at $J=0$\cite{DFisher}, 
\begin{equation} 
P_J(J_{i},\Omega )\approx
\frac{\alpha}{\Omega}(\frac{J_{i}}{\Omega})^{-1+\alpha}~~\Theta (\Omega -
J_{i}).  
\end{equation}
with 
\begin{equation}
\alpha = -\frac{1}{\ln\Omega}
\end{equation}

The fixed form of $P_J(J_{i},\Omega )$
determines the low-temperature behavior of the thermodynamic quantities.
The exponent $\alpha$ gives rise to logarithmic corrections in the
thermodynamic functions.    

The great merit and the reason
for the success of the MDH theory\cite{MDH}, was to show that the
characteristic RS phase  behavior of spin-1/2 chain is universal. More specifically,
it is independent of the original distribution of exchange couplings. The MDH
approach represented a considerable progress with respect to previous
methods\cite{LNBula,GTheo,GTheo2} which required an initial distribution of
bonds with a particular form.  Actually, in order to reproduce the power law
dependence of the low temperature thermodynamic quantities observed
experimentally, the previous models required a initial distribution already
with a power law behavior\cite{GTheo,GTheo2}. 

We turn now to the numerical simulation of chains with spins S=1 and S=3/2.
Each chain is composed of N spin-S objects with periodic boundary conditions
(with N= 50,000). We carried out the averages 
over 10 different configurations for each $P_J(J_{i},\Omega )$ to obtain the
free energy of the system.
We start by choosing the exchange coupling from an uniform distribution
$P_J(J_{i},\Omega )$ given by:
\begin{equation}
P_J(J_{i},\Omega )=\left\{~~\frac{1}{1-\Delta}~~~\hfill\hbox{for $\Delta \leq
J_{i}\leq 1$},     \atop 0~~  \quad  \hbox{otherwise}\right .  
\label{pdel}
\end{equation}
The parameter $\Delta$  represents the strength of the initial disorder of 
the couplings. The case $\Delta=0$  represents the strong disorder limit  
because the distribution becomes extremely broad in a logarithmic scale; 
namely, infinitesimally weak
bonds appear. The weak disorder regime is represented by finite values of $\Delta$.
The distribution, in this case,  presents a gap $\Delta$.

We have studied the flow of the initial coupling distribution,
Eq.~(\ref{pdel}), for the spin-1 chain for the following cases:
$\Delta=0$, $\Delta=0.05$, $\Delta=0.1,$ and $\Delta=0.2$.
We found that for any $\Delta$,  
successive elimination transformations give rise to weaker and weaker
couplings as the cutoff  $\Omega$ decreases, {\it independently of the
temperature}. For sufficiently small  $\Omega$, i.e., after a sufficient number
of eliminations, the distribution  $P_J(J_{i},\Omega )$ becomes peaked at
$J_{i}=0$ and, as in the T=0 case\cite{bia2,Hyman,Jolicoeur,bia3}, can be 
approximated by a power-law. We should point out that the convergence to a
power-law depends on the value of $\Delta$. It is faster in the case of the
gapless ($\Delta =0$) initial distribution.
As in the case of spin-1/2 chains~\cite{MDH} the power-law exponent obtained
for the S=1 chain is non-universal: it depends on
temperature and cutoff, although weakly. In Fig.~\ref{graf1}  we present, for
completeness, the power-law behavior of  $P_J(J_{i},\Omega )$ with exponent
$\alpha = 0.38$ obtained from an initial distribution with $\Delta =0$, at a
fixed temperature $k_{B}T=0.10\Omega$.

Now we focus on the free energy per spin and the specific heat of the spin-1
chain. In  Fig.~\ref{graf2} we show the plot of the specific heat versus
temperature (given in units of $\Omega$). Despite the  power-law
dependence of  $P_J(J_{i},\Omega )$ at low energies described above, we find a
non physical behavior in  the free energy of the system. 
For intermediate temperatures, the free energy gives rise to a negative
specific heat (see Fig.~\ref{graf2}). Nonetheless, at very low temperatures
the free energy and thus the specific heat is well-behaved, being described by
a power-law (see inset of Fig.~\ref{graf2}). These very low temperature
features are similar to the random singlet phase of the disordered spin-1/2
chain \cite{MDH}.  Moreover, at higher temperatures, the free
energy curve  is again well-behaved and gives rise to a physically meaningful
specific heat.  

The presence of the factor $2S(S+1)/3$ in Eq.~(\ref{Kprime}), which defines the
renormalized coupling $J^{\prime}$, plays an important role in our discussion.
In the case of the spin-1 chain it is $4/3$. 
 At low temperatures ($k_{B}T<0.37\Omega$), the function $W_{1}(\beta
\Omega)$ is larger than $3/4$, consequently, when the couplings neighboring
the cutoff $\Omega$ are not small enough, couplings larger than those
eliminated may be generated (see Eqs.~(\ref{Kprime}) and (\ref{Wb})). In
this low temperature regime, although the cutoff decreases  rapidly, large
couplings are generated with non negligible probability. When the couplings
between the strongly coupled pair of spins and its neighbors are not
sufficiently weak, the perturbative scheme fails, giving rise to the spurious
result observed in the free energy. At higher temperatures
($k_{B}T>0.37\Omega$) the MDH method gives correct results for the
thermodynamic  properties~\cite{bia2}. There the function
$W_{1}(\beta \Omega)<3/4$ cancel the coefficient 4/3 of $J^{\prime}_{1}$ (see
Eq.~(\ref{Wb})), thus validating  the
present approach, independently of the starting distribution. 
In addition, when one considers  distributions with a finite gap
(weakly disordered distributions), the breakdown of the MDH formalism is most
pronounced.  The probability of obtaining, at low temperatures, the product 
$J_{1}~J_{3}<3/4\Omega^{2}$  depends on the value of the gap. By increasing
$\Delta$, the initial distribution will not generate
infinitesimally weak couplings to ensure that the product is sufficiently
small. 

In the spin-3/2 chain we performed the decimation process by starting with a
 uniform gapless probability distribution. 
Despite  the larger coefficient in Eq.~(\ref{Kprime}), $P_J(J_{i},\Omega )$
still  develops  a power-law behavior as the cutoff $\Omega$ becomes
sufficiently small. However, this convergence is much slower than that for a
spin-1 chain.  
Now the couplings $J^{\prime}$ which are larger than the decimated out coupling
$\Omega$ are statistically predominant.
The effect of the breakdown of the MDH
method in  the free energy of the $S=3/2$ chain is more dramatic yielding a 
negative specific heat even in the very low temperature limit.

These results has led us to try to improve the MDH transformation in two
directions: first we considered the decimation of larger spin
clusters. In order to gain an  insight on whether or not  this procedure would 
work, it was  sufficient to carry out the decimation at zero
temperature. The intent was to investigate whether the perturbation scheme,
using larger clusters, 
would decrease 
the factor appearing in the expression for the new coupling
$J^{\prime}$.  We have carried out
numerically the decimation process for RAQ chains  with both spin $S=1/2$ and
$S=1$, considering clusters with $2$, $4$, $6$ and $8$ spins. In our numerical
calculations we considered, for simplicity,  $J_i=\Omega =1$, where the $J_i's$
stand for the bonds between neighboring spins in the cluster.  The results for
the case of a cluster with only two spins can be directly  compared with the
value $1/2$ and $4/3$ of the original MDH procedure (see Eq.~(\ref{Kprime})) in
the  S=1/2 and S=1 RAQ chains, respectively. We list the results in
Table~\ref{tab1}. In the case of a spin-1/2 chain the results indicate that the MDH scheme 
is always improved. However, the resulting values for larger clusters in
the spin-1 chain  makes clear that the problem remains unsolved.

We also tried to improve the MDH scheme by considering terms of higher order in
the perturbation expansion, at zero temperature. In
that case, our purpose was  to investigate whether  the higher order
corrections would decrease the factor $2S(S+1)/3$ in Eq.~(\ref{Kprime}).
We have thus calculated  the $T=0$ renormalized coupling, for the spin-1 chain,
up to third order in $\Omega $, which is given by 
\begin{eqnarray}
J'=\frac{4}{3\Omega}J_{1}J_{3}+\frac{1}{\Omega^2}(J^{2}_{1}J_{3}+J_{1}J^{2}_{3})
\label{3ordem}
\end{eqnarray}
Once again, we come upon a result which reinforces that the MDH formalism is
not suitable to describe the low temperature properties of the RAQ chains with
$S\ge 1$ (defined in Eq~(\ref{eq1.1})). Moreover, in the fourth-order
correction to the renormalized bond  biquadratic exchange couplings are
generated. In this case we should consider these interactions already from the
beginning in the Hamiltonian model.  Finally  we call attention that
starting with a power law distribution, the MDH procedure works very well. In
this case, only pairs of spins weakly coupled are present in  the chain and
the probability of the generated coupling  being larger than the one
eliminated is statistically negligible. The singular form of $P_J(J_{i},\Omega
)$ dominates the  thermodynamic properties which are also described by
power-laws.  Of course this is in strong contrast with the robust and
universal character of the RS phase  for the spin-1/2 chain, which is indeed a
new fixed point~\cite{MDH,DFisher}. 

In summary, at zero temperature there is not a safe criterion to discuss the
possibility of the occurrence of the RS phase  in the RAQ chains. In fact, it
was necessary to extend the MDH scheme, for $S\ge 1$, to finite temperature to
establish the failure of the perturbative approach. 
In view of the results we have obtained, we strongly claim that the MDH
procedure is not adequate to study the RAQ chains  with $S\ge 1$ defined in
Eq.~(\ref{eq1.1}). Moreover, the existence of a RS phase at low
temperature is a property assured only for spin-1/2 RAQ chains.

The authors are grateful to the Brazilian agencies FAPERJ and CNPq 
for financial supports.

\newpage

\begin{center}
TABLE I
\end{center}

\begin{table}[hbt]
\begin{center}
\begin{tabular}{|c||c|c|} 
cluster size  &      spin-1/2            &      spin-1      \\ \hline\hline  
2             &       0.500000           &     1.333333         \\ \hline
4             &       0.377992           &     1.881373          \\ \hline
6             &       0.312648           &     2.597776           \\ \hline
8             &       0.270240           &        -             
\end{tabular}
\end{center}
\caption[t1]{Renormalized coupling for chains with $S=1/2$ and $S=1$ obtained
with  clusters of size 2,4,6 and 8. We have taken $J_i=\Omega =1$.}  
\label{tab1}
\end{table}

\begin{center}
FIGURES CAPTION
\end{center}

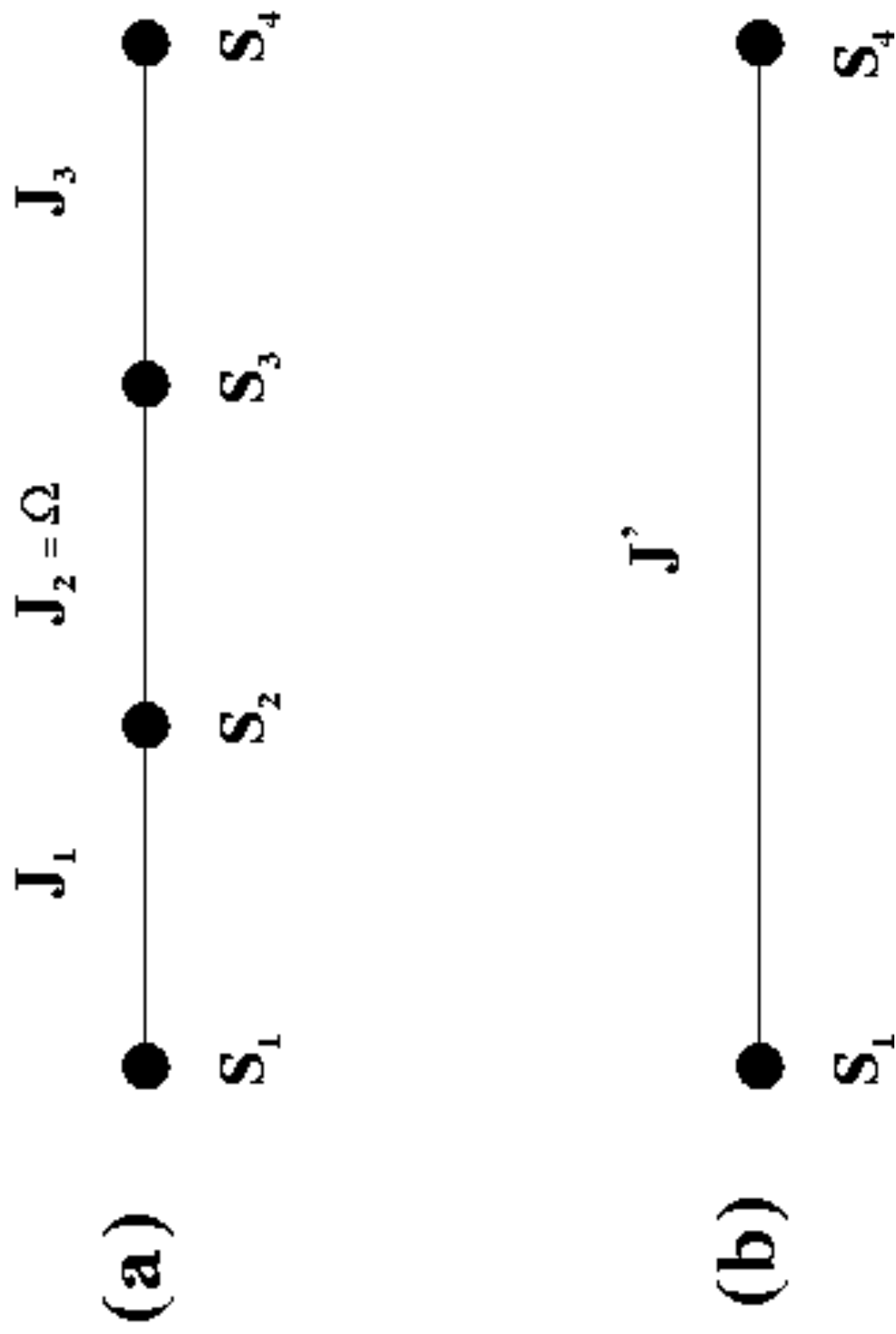
\begin{figure} 
\caption{Spins and coupling constants involved in
the elimination transformation (a) to (b).}
\label{cad} 
\end{figure}
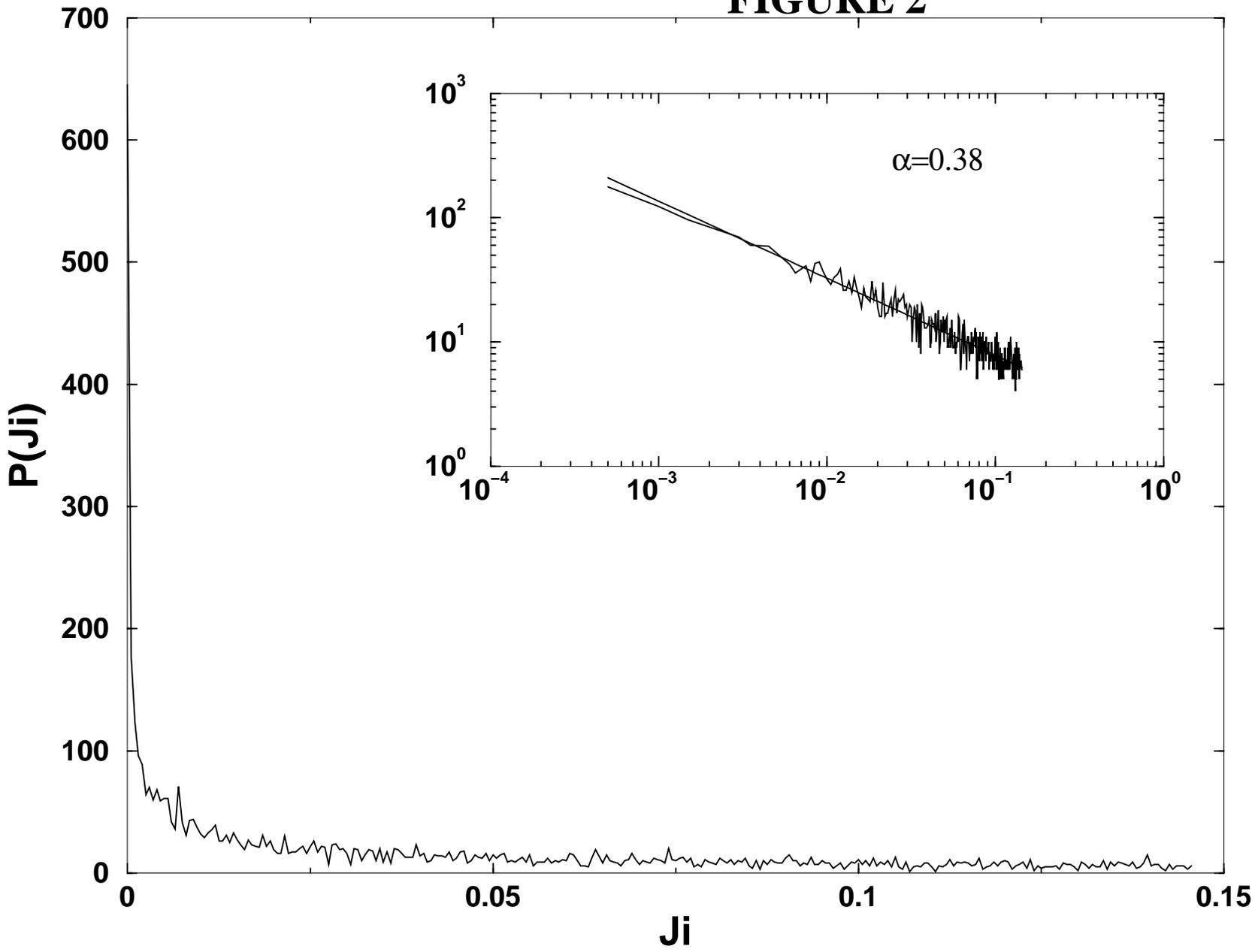
\begin{figure}
\caption{Power-law behavior of 
$P_J(J_{i},\Omega )$  in the low-energy limit, at  $k_{B}T=0.10\Omega$.
Inset: the Log-Log plot of $P_J(J_{i},\Omega )$ yields the exponent $\alpha =
0.38$.} 
\label{graf1} 
\end{figure}
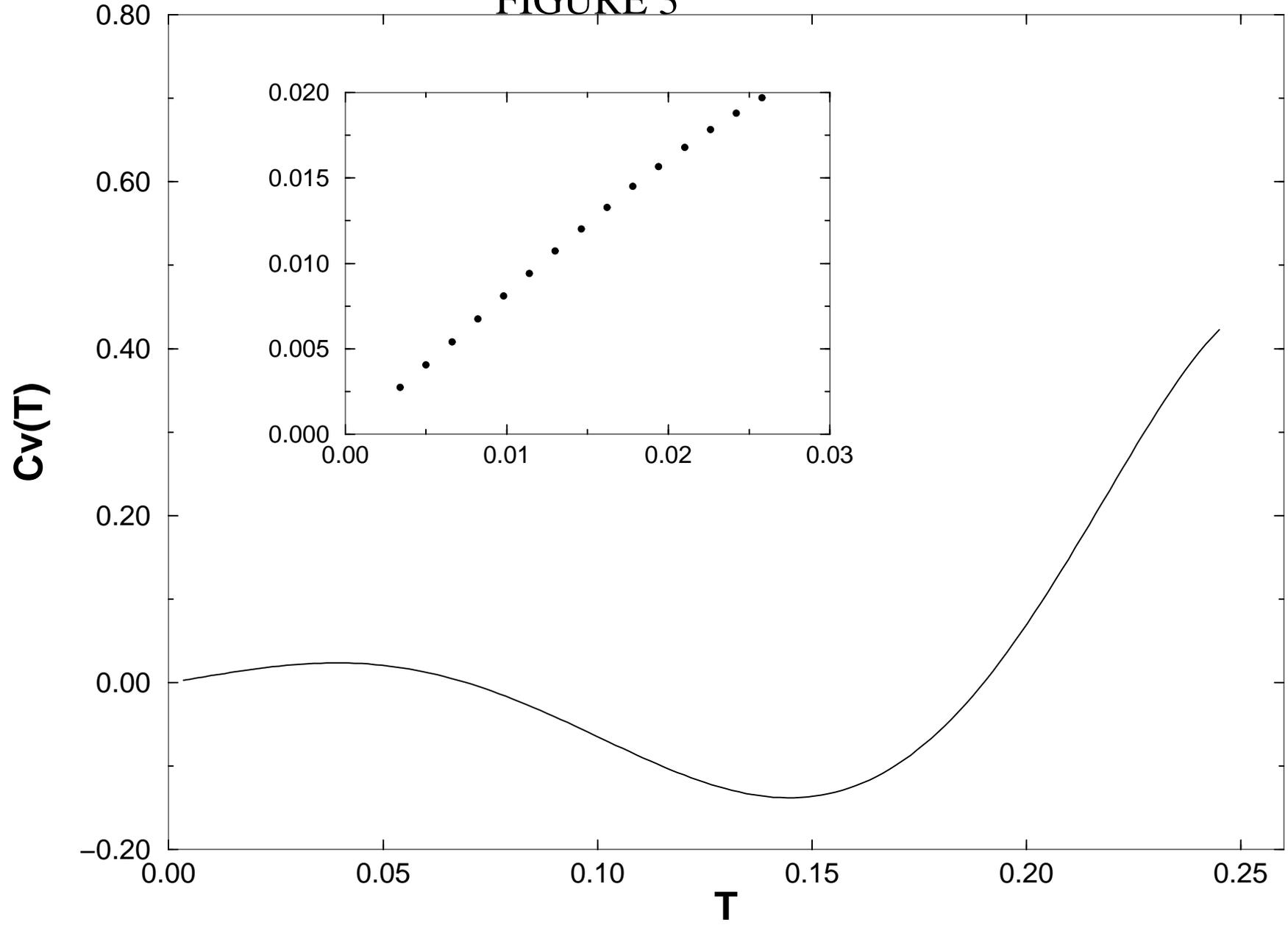
\begin{figure} 
\caption{Low temperature behavior of the specific heat of the random spin-1
chain  for $\Delta =0$. Inset: the power-law behavior with exponent
equal to $0.77$ }.
\label{graf2} 
\end{figure}
\end{document}